%% file: NEWAST-D-10-00147R2.tex
\documentclass[preprint,authoryear,12pt]{elsarticle}

\usepackage{amssymb}
\usepackage{graphicx}
 \usepackage{amsthm}
 \usepackage{astron}
 \usepackage{hyperref}
 \usepackage{breakurl}
\usepackage{setspace}
  \usepackage{rotating}
  \usepackage{tabularx}
  
 \def\astrobj#1{#1}

\journal{New Astronomy}

\begin{document}
\bibliographystyle{astron.bst}
\input{aas_macros.tex}

\begin{frontmatter}


\title{Diffuse Thermal X-Ray Emission in the Core of the Young Massive Cluster Westerlund 1}

\address[label1]{School of Physical Sciences, Dublin City University, Dublin 9, Ireland}
\address[label2]{National Centre for Plasma Science and Technology, DCU, Dublin 9, Ireland}
\address[label3]{School of Cosmic Physics, Dublin Institute for Advanced Studies, Dublin 2, Ireland}

\author[label1,label2]{P. J. Kavanagh}
\ead{patrick.kavanagh8@mail.dcu.ie}

\author[label1,label2]{L. Norci}
\author[label3,label1]{E.J.A. Meurs}


\begin{abstract}
We present an analysis of the diffuse hard X-ray emission in the core of the young massive Galactic cluster Westerlund 1 based on a 48 ks XMM-Newton observation. Chandra results for the diffuse X-ray emission have indicated a soft thermal component together with a hard component that could be either thermal or non-thermal. We seek to resolve this ambiguity regarding the hard component exploiting the higher sensitivity of XMM-Newton to diffuse emission. Our new X-ray spectra from the central (2' radius) diffuse emission are found to exhibit He-like Fe 6.7 keV line emission, demonstrating that the hard emission in the cluster core is predominantly thermal in origin. Potential sources of this hard component are reviewed, namely an unresolved Pre-Main Sequence population, a thermalized cluster wind and Supernova Remnants interacting with stellar winds. We find that the thermalized cluster wind likely contributes the majority of the hard emission with some contribution from the Pre-Main Sequence population. It is unlikely that Supernova Remnants are contributing significantly to the Wd1 diffuse emission at the current epoch.

\end{abstract}


\end{frontmatter}

\section{Introduction}

Since their launch in 1999, the Chandra and XMM-Newton observatories have revolutionized the study of X-ray emission from stellar clusters. Chandra's ACIS and XMM-Newton's EPIC have allowed unprecedented analysis of point sources and diffuse emission in such objects. Of particular importance are the observations of extragalactic Super Star Clusters (SSCs). SSCs\footnote{We note that there exists some ambiguity in the literature as to the classification of SSCs. We follow here one of the many sets of classification criteria, as in \citet{Whitmore2000}} are young (1-10 Myr), massive ($10^{5}-10^{7}$ M$_{\odot}$) objects with extremely dense cores ($\lesssim 10^{5}$ M$_{\odot}$ pc$^{-3}$) and are the predominant sites of massive star formation in starburst and interacting galaxies \citep[eg. \astrobj{NGC 4038/39} and \astrobj{M82},][]{Whitmore1999,Melo2005}. However SSCs are not limited to these extreme environments, with some found in objects such as Blue Compact Dwarfs, non-interacting spirals and Ultra Luminous Infrared Galaxies \citep[eg. \astrobj{Henize 2-10} and \astrobj{PKS 1345+12-C1}:][]{Johnson2000,Larsen1999,Rodriguez2007}. Apart from hosting large numbers of massive stars, SSCs also serve to enrich and energize the local Interstellar Medium (ISM) through a shocked outflowing cluster wind. The cluster wind arises from interacting stellar winds from the massive star population and later from SN ejecta. The enrichment of the local ISM by the cluster wind can potentially drive further star formation in the region. In dwarf starburst galaxies, these winds may be powerful enough to produce a galactic outflow, enriching the Intergalactic Medium (IGM) and potentially killing further star formation in the galaxy. Hence, SSCs provide not only a laboratory for the study of massive stars at various stages of evolution but can also provide vital insights into cluster evolution and star formation on large scales. Unfortunately, given the distance to many of these SSCs and their extremely compact nature it is often impossible to resolve the diffuse emission from the point source emission using Chandra or XMM-Newton. However, it is possible to resolve the diffuse and point source emission in local lower mass analogues. Thus, detailed analysis of such nearer objects can provide key insights to the inner workings of SSCs. \astrobj{Westerlund 1} (Wd1) is one such cluster, which holds the distinction of being the most massive young cluster in the Galaxy.

\par Wd1 was discovered in the early sixties and was initially classified as an open cluster \citep{Westerlund1961}. The cluster suffers from significant reddening \citep[$A_{V} \thickapprox 12.9$ mag, ][]{Piatti1998} and, because of this, only recently detailed photometric and spectroscopic analyses have been performed \citep{Clark2002,Clark2004,Negueruela2005,Clark2005}. \citet{Clark2005} found a rich population of evolved OB stars and, using a standard Kroupa \citep{Kroupa2001} initial mass function (IMF), inferred a cluster mass of $\gtrsim10^{5}$ M$_{\odot}$. This is at the lower limit of the SSC mass range and certainly made Wd1 the most massive cluster in the Galaxy. A more recent deep IR study, however, revises this mass estimate somewhat downwards to $\thickapprox 4.5 \times 10^{4}$ M$_{\odot}$ \citep{Brandner2008}. Although this is still bigger than any other known Galactic cluster, it is slightly smaller than extragalactic SSCs. The same study also revised previous estimates of age and distance to $3.6 \pm 0.7$ Myr and $3.55 \pm 0.17$ kpc respectively, which we adopt for our analysis. \citet{Muno2006}, henceforth MU06, used Chandra data to perform a diffuse emission analysis and found that the emission throughout the cluster is dominated by a hard component. However, they were unable to identify the nature of this emission due to the absence of hard emission lines and discussed both thermal and non-thermal origins for the hard component.

\par In this paper we seek to resolve this issue using the XMM-Newton observational data, given the telescopes' greater sensitivity to diffuse emission. As yet these data have only been used in an analysis of the well known magnetar CXOU J164710.2-455216 in this cluster \citep{Muno22006,Muno2007}. In Section 2 we outline the observational data reduction, before briefly discussing the point source analysis in Section 2.1. We follow by presenting the detailed diffuse emission analysis in Section 2.2. In Section 3 we discuss our results before offering our conclusions in Section 4.

\section{Observations and Analysis}

\begin{figure}[htp]
\begin{center}
\includegraphics[height=7cm, width=14cm]{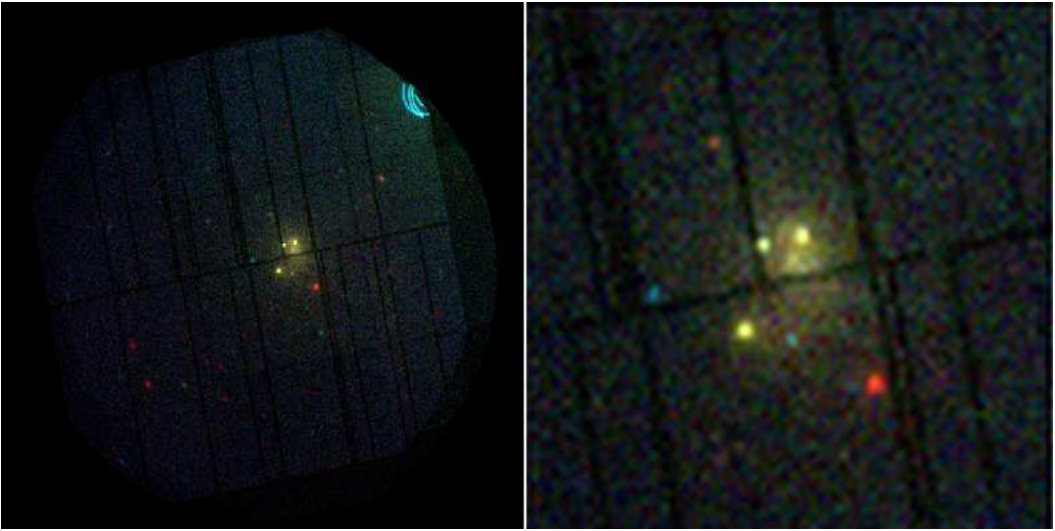}
\caption{Combined MOS/PN false colour images of Westerlund 1 with red, green and blue corresponding to the 0.3-2 keV, 2-4.5 keV and 4.5-10 keV energy bands, respectively. North is up, East is left. Left panel: The entire FOV, approximately 30' in diameter. Both the Wd1 cluster diffuse emission and several point sources are seen at the centre with additional sources scattered throughout the FOV. The X-ray binary 4U 1642-45 reflection artifact is seen in the upper right of the image. Right panel: 5'$\times$5' region centered on Westerlund 1 highlighting the cluster diffuse emission. The bright soft source  to the southwest (seen in red) is the foreground star \astrobj{HD 151018}, an O9Iab star.}
\label{XMM_col}
\end{center}
\end{figure}

XMM-Newton observed Wd1 on 16 September 2006 for $\sim$48 ks (Obs. ID 0404340101, Revolution 1240). The event files were processed using the XMM-Newton Science Analysis Software (SAS, Version 7.1.0)  meta-tasks \textit{emproc} and \textit{epproc}. We then filtered the data for good grades in the energy band 0.3-10 keV (the energy range at which all 3 of the EPIC instruments are most sensitive) and created images for each of the three EPIC cameras, namely the PN, MOS1 and MOS2. The PN and MOS2 images were found to contain single reflection artifacts which are due to X-rays from a source outside the field of view (20'-80' off-axis) reaching the sensitive area of the focal plane detectors by single reflection from the rear end of the hyperboloid component of the XMM-Newton mirror shells\footnote{See \burl{http://xmm.esa.int/external/xmm\_user\_support/documentation/uhb/node23.html}}. This object was identified as the low mass X-ray binary \astrobj{4U 1642-45} which is located approximately 20' to the northwest of the observation aimpoint. Images from the three EPIC instruments were combined to produce the false colour image shown in Figure \ref{XMM_col}.

\subsection{Point Sources}
Point source detection was performed over three standard XMM-Newton energy bands (0.5-2 keV, 2-4.5 keV and 4.5-7.5 keV) on the three EPIC images using the SAS meta-task \textit{edetect-chain}. In total, 90 sources with a minimum maximum-likelihood detection threshold of 10 were found; 7 of these are associated with the reflection and were thus ignored. A further 8 sources were flagged as spurious due to their positions on or near chip gaps and were removed from consideration, leaving 75 source detections in the field. By correlation with the comprehensive Chandra source list in \citet{Clark2008}, 4 of our XMM-Newton sources appear to have high mass stellar X-ray emitting counterparts in the cluster with a further 8 having Pre-Main Sequence (PMS) stellar objects. One other source in the cluster area \citep[within 5' of the cluster centre,][]{Muno2006} was found to have no counterpart in the source list of \citet{Clark2008} or in the SIMBAD database and is likely a newly detected flaring PMS star. Table \ref{pointsources} gives our detected cluster sources and their corresponding Chandra designations, along with spectrally derived source parameters.

\begin{sidewaystable}
\begin{minipage}[t]{\textwidth}
\begin{center}
\caption{Westerlund 1 XMM-Newton Point Sources.}
\begin{scriptsize}
\label{pointsources}
\renewcommand{\footnoterule}{}  
\begin{tabular}{ccccccccccccc}
\hline
\multicolumn{13}{c}{High Mass Stars} \\
\hline \hline
No. & Chandra Source & Opt. ID & Sp Type & X-ray Type & MOS Net Counts & Model & $N_{H}$ & $kT$ & $kT_{2}$ & $\chi^{2}/\nu$ & $F_{\mathrm{X}}^{\mathrm{unabs}}$ & $L_{\mathrm{X}}^{\mathrm{unabs}}$\\ 
 & (\begin{tiny}CXO J\end{tiny}) & & & & & & ($10^{22}cm^{-2}$) & ($keV$) & ($keV$) & & (\begin{tiny}$10^{-12}$ erg cm$^{-2}$ s$^{-1}$\end{tiny}) & (\begin{tiny}$10^{33}$ erg\ s$^{-1}$\end{tiny}) \\
 &  &  &  & \textit{(a)}  & \textit{(b)}  & \textit{(c)}  &  &  &  &  & \textit{(d)} & \textit{(d)} \\
  
\hline \\

 & 164704.1-455039 & W 30 & \textit{O9-B0.5Ia} & \textit{CWB} & & & & & & \\

1 & 164704.1-455031 & W 9 & \textit{sgB[e]} & \textit{CWB} & 2157 & 2T & $2.5$ & $0.7$ & $2.9$ & 2.170 & 4.37 & 6.59 \\

 & 164705.1-455041 & W 27 & \textit{OB SG} & \textit{RDIS} & & & & & & \\
 & & & & & & & & & & & & \\

2 & 164705.0-455225 & WR F & \textit{WC9d} & \textit{CWB} & 509 & 1T & $0.77^{1.26}_{0.29}$ & $4.72^{12.90}_{2.93}$ & -- & 0.909 & 0.23 & 0.26 \\
3 & 164708.3-455045 & WR A & \textit{WN7b} & \textit{CWB} & 988 & 1T & $2.37^{2.85}_{2.05}$ & $1.48^{1.78}_{1.17}$ & -- & 1.967 & 0.89 & 1.34\\ 
4 & 164710.2-455217 & -- & -- & \textit{Magnetar} & 1708 & BB & $0.90^{1.07}_{0.76}$ & $0.63^{0.66}_{0.60}$ & -- & 1.084 & 0.36 & 0.54 \\ 
\\*
\hline
\multicolumn{13}{c}{Pre-Main Sequence Stars} \\
\hline \hline
No. & Chandra Source & Opt. ID & Sp Type & X-ray Type & MOS Net Counts & Model & $N_{H}$ & $kT$ & $kT_{2}$ & $\chi^{2}/\nu$ & $F_{\mathrm{X}}^{\mathrm{unabs}}$ & $L_{\mathrm{X}}^{\mathrm{unabs}}$\\ 
 & (\begin{tiny}CXO J\end{tiny}) & & & & & & ($10^{22}cm^{-2}$) & ($keV$) & ($keV$) & & (\begin{tiny}$10^{-12}$ erg cm$^{-2}$ s$^{-1}$\end{tiny}) & (\begin{tiny}$10^{33}$ erg\ s$^{-1}$\end{tiny}) \\
 &  &  &  & \textit{(a)}  & \textit{(b)}  & \textit{(c)}  &  &  &  &  & \textit{(d)} & \textit{(d)} \\
  
\hline \\
5 & 164640.8-454834 & -- & -- & \textit{PMS Flare} & 34 & -- & -- & -- & -- & -- & -- & -- \\

6 & 164648.8-455307 & -- & -- & \textit{PMS Flare} & 39 & 1T & $1.36^{2.70}_{0.55}$ & $0.86^{2.71}_{0.25}$ & -- & 0.959 & 0.04 & 0.05 \\

7 & -- & -- & -- & \textit{PMS Flare} & 62 & 1T & $0.76^{1.36}_{0.50}$ & $4.07^{7.08}_{2.18}$ & -- & 1.992 & 0.01 & 0.02 \\
 
8 & 164652.6-455357 & -- & -- & \textit{PMS Flare} & 210 & 1T & $0.51^{0.84}_{0.32}$ & $7.64^{--}_{3.22}$ & -- & 1.057 & 0.06 & 0.09 \\

9 & 164703.2-455157 & -- & -- & \textit{PMS Flare} & 397 & 1T & $1.28^{1.96}_{0.65}$ & $2.49^{4.90}_{1.69}$ & -- & 0.954 & 0.12 & 0.19 \\

10 & 164712.8-455435 & -- & -- & \textit{PMS Flare} & 40 & -- & -- & -- & -- & -- & -- & -- \\

11 & 164713.6-454857 & -- & -- & \textit{PMS Flare} & 62 & -- & -- & -- & -- & -- & -- & -- \\

12 & 164718.7-454758 & -- & -- & \textit{PMS Flare} & 66 & -- & -- & -- & -- & -- & -- & -- \\

13 & 164720.1-455138 & -- & -- & \textit{PMS Flare} & 19 & -- & -- & -- & -- & -- & -- & -- \\
\hline \\
\end{tabular}
\end{scriptsize}
\end{center}
\begin{scriptsize}
\begin{singlespacing}
\noindent In this table the Chandra designation, optical ID and spectral types are adopted from \citet{Clark2008} and references therein. Notice that Source 1 encompasses W 30, W 9 and W 27 which are unresolved by XMM-Newton. No counterpart for Source 7 was found in the analysis of \citet{Clark2008} or in the SIMBAD database.\\
\textit{(a)} - X-ray type adopted from \citet{Clark2008}. PMS = Pre-Main Sequence, CWB = Colliding Wind Binary, RDIS = Radiatively Driven Instability Shocks in stellar winds. \textit{(b)} - Combined MOS net counts. PN counts omitted due to some sources lying on detector gaps. \textit{(c)} - Best fit X-ray spectral model each of which was convolved with an absorbing hydrogen column. Only those sources with $>100$ combined MOS and PN net counts were spectrally analysed. All spectral fits were performed in XSPEC 12.3.1. 1T = one temperature thermal plasma, 2T = two temperature thermal plasma, BB = black body. The spectrum of Source 2 is approximated by a 2T thermal plasma model however the true spectrum is likely described by a more complex model given it is an amalgam of several unresolved bright sources. In addition, Sources 3 and 4 are likely described by a 2T thermal plasma model however the soft components are poorly constrained and thus only the 1T approximation is given. \textit{(d)} - Absorption corrected X-ray fluxes and luminosities quoted in the 0.5-8 keV energy range.
\end{singlespacing}
\end{scriptsize}
\end{minipage}
\end{sidewaystable}

\subsection{Diffuse Analysis}
\label{diff-anal}
\begin{figure}[ht]
\begin{center}
\includegraphics[width=7cm, height=7cm]{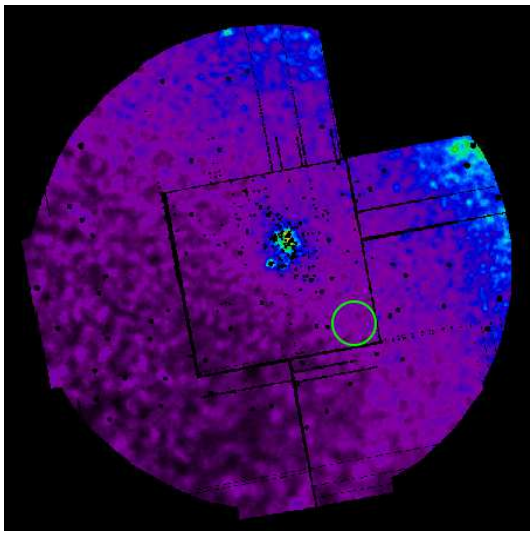}
\caption{Smoothed, non-background subtracted, combined MOS1/MOS2 image in which sources are masked with the CCDs most affected by the reflection omitted. Again the FOV is approximately 30' in diameter. The green circle indicates the region chosen for background extraction. This source free region lies on the same CCD as the cluster core in both the MOS1 and MOS2 images, is outside the total cluster region and is approximately as contaminated by the reflection as the cluster core. A zoomed image of the centre region in this figure can be found in Figure \ref{dist}.}
\label{com_mos_smo}
\end{center}
\end{figure}

It is obvious from Figure \ref{XMM_col} not only that the reflections could contaminate the diffuse emission in Wd1 but also that they are more prominent in the harder energies which are of particular interest to our analysis. To address this we considered various analysis techniques including the XMM-Newton Extended Source Analysis Software (ESAS) and `blank sky' background event files but found that none could adequately account for the reflection emission. Instead we opt for the more traditional method of background extraction from regions within the FOV. By defining background regions that are as contaminated by the reflection as the cluster, the contribution of the reflection to the cluster spectra can be reduced. We decided against using the PN data for the diffuse emission analysis as several detector gaps mask some of the Wd1 cluster core. Therefore, the following analysis is based on the MOS data only. To assess first the diffuse emission in the FOV of the MOS cameras we create a non-background subtracted image of the emission using ESAS (Version 2)\footnote{Software available  at http://xmm2.esac.esa.int/external/xmm\_sw\_cal/ \\background/epic\_esas.shtml}. To exclude the point sources from the image we use an adapted form of the \textit{cheese} task to mask not only the point sources detected in our XMM-Newton analysis but also those detected by \citet{Clark2008}\footnote{Electronic catalogue accessed via the Vizier service} but undetected in our XMM-Newton analysis. For the detected sources we determined appropriate exclusion regions by assessing the radial brightness profile of each source in two energy bands (namely 2-4.5 keV and 4.5-10 keV, since we will focus on the harder energies in the analysis below). In each energy band the extent of the exclusion regions is set by the source brightness profiles approaching local background levels and the largest of these two areas is adopted as the overall source exclusion region. For each of the XMM-Newton sources in the Wd1 cluster these regions are $\sim$30'' in diameter. This is not a perfect solution however since incidental photons from extreme edges of the wings of the source PSF may not be masked. We estimate from the background subtracted source brightness profiles that no more that 10\% of the photon flux could have been missed by these exclusions regions. No such method can be employed to mask the undetected sources due to the simple unavailability of the source brightness profiles. Because of this, 10'' diameter regions were used to mask as much as possible any detected photons from these sources (for which the greatest probability for these photons is to be found in the inner PSF core) without sacrificing large numbers of diffuse source photons. The resulting image is shown in Figure \ref{com_mos_smo}.

\par Figure \ref{com_mos_smo} clearly shows that the diffuse emission in the FOV is dominated by the reflection. Extracting background spectra from within the FOV would mean that any contributing sources of background photons will be contained in these spectra, including the reflection itself. Hence, we first assess each contaminant as some of these can vary across the FOV. The ESAS manual identifies 4 sources of background contamination. These are Solar Wind Charge Exchange (SWCE) contamination, instrumental fluorescence lines, residual soft proton contamination and the cosmic background.  In the case of this observation, the SWCE and one instrumental fluorescence line can largely be ignored as they only significantly contaminate spectra below 1.5 keV (given the absorbing hydrogen column in the direction of Wd1, any emission below this energy is likely foreground emission). The second fluorescence line ($E = 1.75$ keV) may contaminate the spectra especially in larger extraction regions. Soft proton contamination blights almost all observations. Periods of high contamination are screened out in the data reduction process but residual contamination can still affect the observational data. However, it is assumed that the level of contamination is constant across the FOV and hence should be contained in the backgrounds. Similarly it is assumed that the cosmic background is constant across the FOV and is also contained in the backgrounds. The final source of contamination, unique to this observation, is the 4U 1642-45 reflection. This can be compensated for by the selection of appropriate background regions that are equally contaminated as the cluster, determined using Figure \ref{com_mos_smo}.

\par
\begin{figure}[ht]
\begin{center}
\includegraphics[height=7.5cm, width=7.5cm]{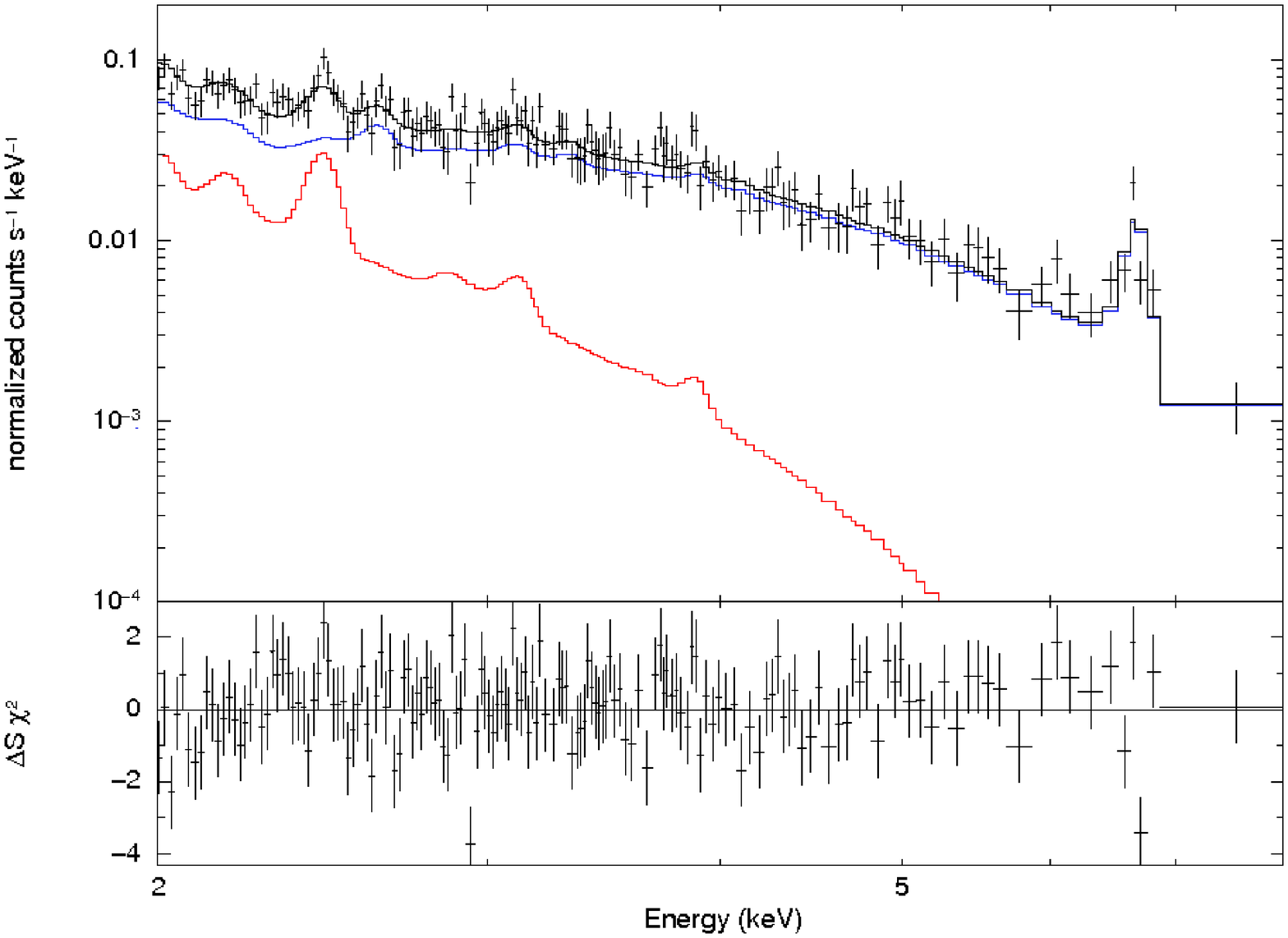}
\caption{Combined MOS inner 2' diffuse emission spectrum binned with S/N=3 and fit with an absorbed two temperature thermal plasma model. The blue and red lines indicate the individual thermal components of the model, representing the hard and soft components respectively.}
\label{line}
\end{center}
\end{figure}

 We adopt the cluster core coordinates determined by MU06 as $\alpha_{0}$ = 16 47 04.3, $\delta_{0}$ = -45 50 59 and extract spectra from the central 1' radius region and three annuli extending out from the core (1'-2', 2'-3.5', 3.5'-5') with the XMM-Newton and Chandra point sources masked. As any X-ray photons detected below 1.5 keV are likely foreground emission and because of the two instrumental fluorescence lines at 1.49 keV and 1.75 keV, we have restricted our spectral analysis to the 2-8 keV energy band which conforms to the standard Chandra hard band as used by MU06. The MOS 1 and MOS 2 spectra and ancillary files for each annulus were combined\footnote{Following the procedure found at \burl{http://xmm.esa.int/sas/current/documentation/threads/epic_merging.shtml}} and the resulting spectra were adaptively binned so that each bin has a S/N of 3. MU06 found that the cluster diffuse emission spectra are well fit with either an absorbed two-temperature thermal plasma model (the harder thermal component with sub-solar abundance to explain the lack of hard emission lines) or an absorbed thermal plasma plus power law model. Hence, the combined MOS spectra were fit in XSPEC\footnote{XSPEC Version 12.3.1 was used for all spectral fits in this analysis.} with these models. The abundance parameter of the cool thermal component in the fits was poorly constrained due to the majority of this emission falling below 1.5 keV, however it was fixed at 2 $\mathrm{Z}_{\odot}$ to be consistent with MU06. The fit results are in general agreement with those of MU06. The outer annulus exhibits only a hard component adequately fit by either an absorbed thermal plasma or absorbed power law model. The best-fit absorbed two temperature thermal plasma and absorbed thermal plasma plus power law models yield very similar best-fit statistics in the inner annuli due to the absence of hard emission lines, as also found by MU06. This is somewhat surprising as, for a cluster such as Wd1 with a large, centrally located WR population, the hard diffuse emission is expected to be thermal in origin due mostly to a thermalized cluster wind \citep{Oskinova2005}. To assess this further we extracted and combined MOS spectra from the inner 2' radius region which again was adaptively binned so that each bin has a S/N of 3. The resulting combined MOS spectrum is shown in Figure \ref{line}.

\begin{center}
\begin{sidewaystable*}
\begin{minipage}[t]{\textwidth}
\caption{Diffuse Spectral Fits }
\begin{footnotesize}
\label{diffuseparam}
\centering
\renewcommand{\footnoterule}{}  
\begin{tabular}{cccccccccc}
\hline \hline
\multicolumn{9}{c}{Two temperature thermal plasma} \\

Region & $N_{\mathrm{H}}$ & $kT_{1}$ & $(Z/\mathrm{Z}_{\odot})_{1}$\footnote{$(Z/\mathrm{Z}_{\odot})_{1}$ for both models is fixed at 2.} & $kT_{2}$ & $(Z/\mathrm{Z}_{\odot})_{2}$ & $\chi^{2}/\nu$ & $F_{\mathrm{X}}^{\mathrm{unabs}}$ \footnote{Absorption corrected X-ray fluxes and luminosities quoted in the 2-8 keV energy range, consistent with the analysis of MU06.} & $L_{\mathrm{X}}^{\mathrm{unabs}}$ $^{b}$ & $L_{\mathrm{X, SB}}^{\mathrm{unabs}}$ \footnote{Diffuse emission surface brightness in the 2-8 keV energy range}\\ 
 & (\begin{tiny}$10^{22}$ cm$^{-2}$\end{tiny}) & (\begin{tiny}keV\end{tiny}) &  & (\begin{tiny}keV\end{tiny}) & &  & (\begin{tiny}$10^{-12}$ erg cm$^{-2}$ s$^{-1}$\end{tiny}) & (\begin{tiny}$10^{33}$ erg\ s$^{-1}$\end{tiny}) & (\begin{tiny}$10^{33}$ erg\ s$^{-1}$\ pc$^{-2}$\end{tiny}) \\ 
\hline \\

$<$2' & $2.03^{2.14}_{1.88}$ & $0.68^{0.80}_{0.55}$ & 2 & $3.07^{3.67}_{2.69}$ & $0.62^{0.89}_{0.40}$ & $0.971$ & $1.71$ & $2.56$ & $0.27$\\ 
\\*
\hline \\

\multicolumn{9}{c}{Thermal plasma plus power law} \\

Region & $N_{\mathrm{H}}$ & $kT_{1}$ & $(Z/\mathrm{Z}_{\odot})_{1}$ $^{a}$ & $\Gamma$ & - & $\chi^{2}/\nu$ & $F_{\mathrm{X}}^{\mathrm{unabs}}$ $^{b}$ & $L_{\mathrm{X}}^{\mathrm{unabs}}$ $^{b}$& $L_{\mathrm{X, SB}}^{\mathrm{unabs}}$ $^{c}$ \\ 
 & (\begin{tiny}$10^{22}$ cm$^{-2}$\end{tiny}) & (\begin{tiny}keV\end{tiny}) &  &  & &  & (\begin{tiny}$10^{-12}$ erg cm$^{-2}$ s$^{-1}$\end{tiny}) & (\begin{tiny}$10^{33}$ erg\ s$^{-1}$\end{tiny})  & (\begin{tiny}$10^{33}$ erg\ s$^{-1}$\ pc$^{-2}$\end{tiny}) \\ 
\hline \\

$<$2' & $2.07^{2.32}_{1.81}$ & $0.81^{0.97}_{0.72}$ & 2 & $2.43^{2.62}_{2.21}$ & - & $1.142$ & $1.50$ & $2.27$ & $0.24$\\ 

\end{tabular}
\end{footnotesize}
\end{minipage}
\end{sidewaystable*}
\end{center}

With this choice of extraction region, the He-like Fe 6.7 keV line, which is used as a diagnostic for thermal diffuse hard emission, is now seen in the combined MOS spectrum.  A straightforward summation of the line flux above continuum, combined with the counts statistics, yields a line significance of $6.2\ \sigma$.  We fit the 2-8 keV combined core MOS spectrum in XSPEC with various combinations of thermal and non-thermal models and find that the data are best fit with an absorbed two temperature thermal plasma model as this can account for the He-like Fe 6.7 keV line and the softer emission lines below 3 keV due to He-like S and He-like Si. The results of the core two temperature thermal plasma and, for comparison, the thermal plasma plus power law fits are given in Table \ref{diffuseparam}, together with values for the diffuse X-ray surface brightness corresponding to these fits. The fact that the two temperature thermal plasma model is a better fit as only this can account for the 6.7 keV emission line demonstrates that, at least in the inner 2' radius region, the hard component is predominantly thermal in origin. We applied a similar treatment to both the outer extraction annuli, however this failed to reveal any line emission in the hard continuum so it is still debatable as to whether thermal or non-thermal processes are responsible for the hard emission in these regions.
Diffuse non-thermal X-ray emission has been found for only a few star formation
regions in the Galaxy and in the LMC \citep[][and references therein]{Maddox2009}. As a
successful fit was obtained for the Wd1 diffuse emission including the Fe 6.7 keV (Table
\ref{diffuseparam}), this leaves little room for any additional non-thermal contribution. Nevertheless, we have run several further, explorative spectral fitting attempts with XSPEC for models with
thermal plus non-thermal components. However, even when fixing non-thermal fitting
parameters to reasonable choices, no convergence of the fits could be obtained. Thus, no
evidence for an additional diffuse non-thermal emission component was found for Wd1.


\section{Discussion}

\subsection{Origin of 6.7 keV Emission Line}
To confirm that the emission line at 6.7 keV is a feature of the cluster diffuse emission we must rule out other potential sources for this line, such as point source contamination, the reflection, cosmological background and those other contributions mentioned above.

\subsubsection{Point Source Contamination}
\begin{figure}[ht]
\begin{center}
\includegraphics[height=7.5cm, width=7.5cm]{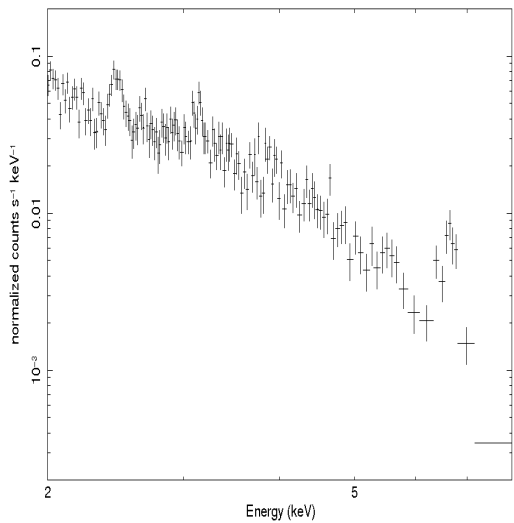}
\caption{Combined MOS spectrum of point sources within a 2' region centered on the cluster core.}
\label{comp}
\end{center}
\end{figure}

Although exclusion regions were defined to mask the point sources it is possible that these masks did not completely exclude point source photons at the extreme edges of the wings of their PSFs. To verify that the emission line is not the result of point source contamination, combined point source spectra for those sources that lie within the 2' radius circle centered on the cluster core were extracted for each MOS camera. The resulting spectra were combined and this spectrum adaptively binned so that each bin has a S/N of 3, shown in Figure \ref{comp}.

Inspection of the 6.7 keV region of the combined MOS point source spectrum in Figure \ref{comp} shows line emission centered at $\sim 6.7$ keV with a significance of 7.3 $\sigma$. To assess the strength of this line in comparison to the 6.7 keV line in the diffuse emission spectrum (which excludes the point sources) we determine the X-ray flux in the 6.6 keV - 6.8 keV range for the combined MOS diffuse spectrum and the combined MOS point source spectrum. We find that the combined MOS point source spectral line flux of $5.73 \times 10^{-14}$ erg cm$^{-2}$ s$^{-1}$ is just over half the strength of the combined MOS diffuse emission line flux of $1.02 \times 10^{-13}$ erg cm$^{-2}$ s$^{-1}$. If the emission line in the diffuse spectrum were a result of point source contamination we would expect the line flux in the point source spectrum to be significantly larger than that in the diffuse spectrum given that the majority of these photons was in fact masked out when creating the diffuse spectrum. The exclusion regions for the sources detected in the XMM-Newton observation were defined based on their brightness profiles in the 2-4.5 keV and 4.5-10 keV energy bands (see Section \ref{diff-anal}) with the largest determined region from these bands set as the overall exclusion region for the respective sources. The exclusion regions were invariably determined by the 2-4.5 keV brightness profiles given the higher number of photons in this energy band in comparison to the 4.5-10 keV band for each of the sources. In addition it was determined that no more than 10\% of the source photon flux is missed by these extraction regions. Since in this discussion we are focussing on the 6.6-6.8 keV energy range and given that for sources with off-axis angle $<$ 2' in EPIC MOS observations (which is our Wd1 core radius) the fractional encircled energy radii decrease with increasing photon energy\footnote{See \burl{http://xmm.esa.int/external/xmm\_user\_support/documentation/uhb/node18.html}}, the exclusion radii for the sources should be more than sufficient to mask $>$90\% of the source photons in the 6.6-6.8 keV energy range. As such we expect the point source contamination due to photons from the extreme wings of their PSFs to contribute $<5$\% of the observed diffuse emission line flux. Additionally, the point source extraction regions occupy $\sim$25\% of the cluster core area which is one third of the diffuse emission extraction region. As such one third of the combined diffuse emission flux in the 6.6 keV - 6.8 keV range is likely contained in the combined point source spectral line flux, meaning the flux due to point sources only may be as low as $2.33 \times 10^{-14}$ erg cm$^{-2}$ s$^{-1}$ and contribute  $<<$5\% of the observed diffuse line flux. For these reasons we can rule out point source contamination.

\subsubsection{Reflection}
\label{reflection}
\begin{figure}[ht]
\begin{center}
\includegraphics[height=7.5cm, width=7.5cm]{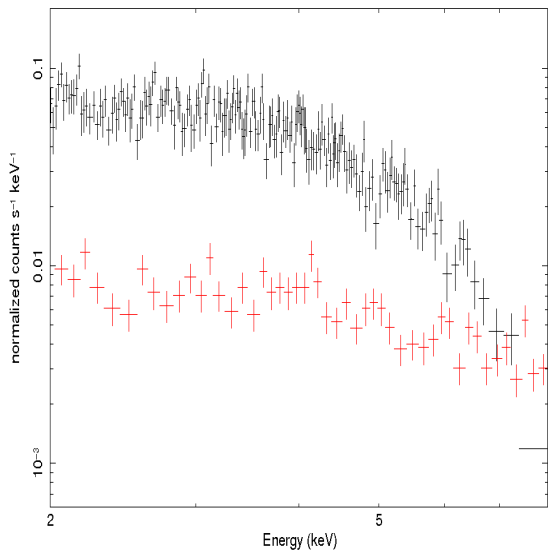}
\caption{The combined MOS background-subtracted reflection spectrum, extracted from those regions most contaminated by the strong NW out-of-FOV source, is shown in black. The background spectrum, shown in red, is extracted from regions devoid of point sources on the opposite (SE) side of the FOV, where the contamination from the reflection is at a minimum. The background spectrum was binned so that each bin contains 30 counts.}
\label{ref}
\end{center}
\end{figure}

To reduce the contribution of the reflection to the diffuse spectra we selected background regions as much contaminated by the reflection as the cluster. However, this is not a perfect solution and contamination by the reflection has to be investigated. To assess if contamination by the reflection is the source of the emission line, we extract and combine spectra for those regions most affected by the reflection (but without the reflection hyperbolas). We create a new combined MOS background spectrum from regions devoid of point sources on the opposite side of the FOV where the contamination from the reflection is at a minimum (i.e. - mainly the regular background contaminants present) and subtract this from the combined MOS reflection spectrum. The resulting spectrum was again binned so that each bin has a S/N of 3, shown in Figure \ref{ref}.

Figure \ref{ref} shows no significant emission at or around 6.7 keV so we can safely assume that contamination by the reflection is not responsible for the line.

\subsubsection{Instrumental Background}
The instrumental background comprises the soft proton background, instrumental fluorescence lines and particle background. From the ESAS method, it is seen that the soft proton contamination is well modeled as a power law (or broken power law). Hence soft proton contamination cannot be responsible for the emission line. Similarly the instrumental fluorescence lines are modeled as Gaussian peaks at 1.49 keV and 1.75 keV and thus are not the source of the line at 6.7 keV. Moreover, the particle background contribution is simply too low to cause the emission line \citep{Kuntz2008}.

\subsubsection{Cosmic Background}
When extracting the backgrounds we assume that the cosmic background emission is constant across the FOV and thus contained in the backgrounds. To assess the contribution of the cosmic background we use the combined MOS background extracted for use with the reflection spectrum in Section \ref{reflection}, indicated by the red spectrum in Figure \ref{ref}. We see no significant emission lines at or around the 6.7 keV region so we can assume that the diffuse emission line is not from the cosmic background.

\subsection{Sources of Hard Emission}
Having demonstrated that the 6.7 keV line is a feature of the cluster diffuse emission and hence that the diffuse emission within 2' of the cluster core is mostly thermal in origin, we now address possible sources for this component. There are three potential sources of thermal emission in the core of a cluster as large as and with the age of Wd1. These are unresolved PMS stars, a thermalized cluster wind and Supernova Remnants (SNRs).

\subsubsection{Unresolved Pre-Main Sequence Stars}
\label{upms}
In their Chandra point source analysis, \citet{Clark2008} identified not only the X-ray emitting high mass population but also a number of PMS stars down to a limiting luminosity of $\sim10^{31} \mathrm{erg\ s}^{-1}$. These sources represent the most luminous members of the PMS population and were masked in the spectral analysis above. However, the remaining unresolved PMS population could contribute significantly to the diffuse emission in the cluster. To obtain an estimate for the contribution of unresolved PMS population, a similar method to that adopted by \citet{Getman2006,Wang2007,Broos12007,Ezoe2006}; and others was used. 

\begin{figure}[ht]
\begin{center}
\includegraphics[height=7.5cm, width=9cm]{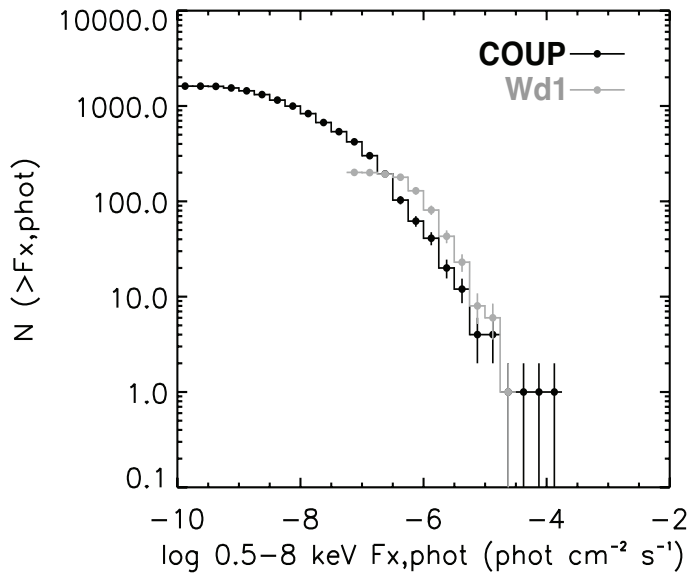}
\caption{Photon flux distributions for the Wd1 and ONC. The ONC distribution has been corrected for the increased distance, foreground absorption and age of Wd1. The errorbars indicate the $1\sigma$ Poisson error.}
\label{flux-dists}
\end{center}
\end{figure}

\noindent We constructed a PMS source 0.5-8 keV photon flux distribution ($F_{\mathrm{X,phot}}$, in units of phot cm$^{-2}$ s$^{-1}$) from the catalogue of \citet{Clark2008} and compared it to that of the Orion Nebular Cluster (ONC) using data from the Chandra Orion Ultradeep Project \citep[COUP, ][]{Getman2005}. We assume that the Wd1 cluster and ONC have the same Initial Mass Function (IMF) and thus X-ray flux distribution, differing only in the size of their underlying populations. However, prior to this, several corrections were applied to  COUP photon fluxes to account for the increased distance, foreground absorption and age of Wd1. To account for the difference in distance the COUP fluxes were simply scaled according to the ratio of the cluster distances. Since the ONC and Wd1 are subject to quite different absorbing hydrogen column densities ($3\times10^{21}$ cm$^{-2}$ \citep{Feig2005} and $\sim2\times10^{22}$ cm$^{-2}$, respectively) we must correct the COUP photon fluxes for the increased absorption to Wd1. We do this by assuming that the composite spectrum of the Wd1 PMS population is the same as that of the COUP, namely a two-temperature thermal plasma with $kT_{1}\ =\ 0.5$ keV and $kT_{2}\ =\ 3.3$ keV \citep{Feig2005}. Using the PIMMS\footnote{See \burl{http://heasarc.nasa.gov/docs/software/tools/pimms.html}} tool we determine an absorption correction for the composite spectrum and apply it to the COUP photon flux distribution. Finally, since the X-ray luminosity of PMS stars has been found to decrease with time $\tau$ as $L_{\mathrm{X}}\ \propto\ \tau^{-0.3}$ \citep{Preib2005}, the COUP fluxes were adjusted to account for the decreased X-ray luminosity of the older PMS population in Wd1. At this stage, the COUP photon flux distribution has been adjusted so that the ONC population effectively has the same age, distance and absorption as Wd1. The photon flux distributions can now be reliably compared, shown in Figure \ref{flux-dists}.

\noindent Figure \ref{flux-dists} indicates that the higher end of the flux distribution of each of the clusters is very similar, differing only in the number of sources per bin, which shows that the earlier assumption of a similar underlying IMF, and hence X-ray flux distribution, was reasonable. The break in the Wd1 distribution just below log $F_{\mathrm{X,phot}}$ $\sim-6$ phot cm$^{-2}$ s$^{-1}$ is due to the incompleteness of the stellar population because of observational constraints. To determine the contribution of the undetected population we scale the COUP distribution to that of Wd1 to determine the overall photon flux of the entire population and subtract that of the detected population, leaving a photon flux due to the unresolved sources of $4\times10^{4}$ phot cm$^{-2}$ s$^{-1}$. We can convert this to a luminosity if we consider that the overall absorption corrected X-ray luminosity of the COUP population responsible for the photon flux distribution is known (overall 2-8 keV $L_{\mathrm{X}}^{\mathrm{unabs}}\sim1.2 \times 10^{33}$ erg\ s$^{-1}$, corrected for the age of Wd1). Scaling this to the unresolved Wd1 population yields a 2-8 keV $L_{\mathrm{X}}^{\mathrm{unabs}}$ of $\sim1.3 \times 10^{33}$ erg\ s$^{-1}$. However, not all of the PMS population of \citet{Clark2008} is located in the core of the cluster but rather spread through the entire cluster area. Since the vast majority is within 3 pc of the cluster centre, we assume that all of the determined unresolved PMS X-ray emission comes from this region, resulting in a surface brightness value of $\sim4 \times 10^{31}$ erg s$^{-1}$ pc$^{-2}$. This is approximately 15\% of the surface brightness of the diffuse emission in the core (see Table \ref{diffuseparam}), meaning an unresolved population can only account for a small fraction of the diffuse emission in the core.

\subsubsection{Cluster Wind}
\label{tcw}
A cluster as large as Wd1 is expected, and is indeed found by \citet{Clark2005}, to contain many massive stars in the cluster core. These massive stars are the source of large amounts of energy and mass being ejected into the cluster volume via stellar winds. The winds collide and thermalize, filling the cluster core volume with a hot, shocked plasma. After some time, the outflow from these thermalized winds becomes stationary and a steady state cluster wind ensues \citep[][and references therein]{Canto2000}. The temperature of this hot, diffuse plasma throughout the core is sufficiently high to radiate at X-ray energies and hence is potentially responsible for the diffuse hard emission in Wd1. MU06 make use of the equations of \citet{Canto2000} for the central hydrogen density and temperature, namely:

\begin{equation}
\left(\frac{n_{0}}{\mathrm{cm}^{-3}}\right) = 0.1 N \left( \frac{\dot{M}}{10^{-5} \mathrm{M_{\odot}\ yr}^{-1}}\right) \left( \frac{v_{\mathrm{w}}}{10^{3} \mathrm{km\ s}^{-1}}\right)^{-1} \left( \frac{R_{\mathrm{C}}}{\mathrm{pc}}\right)^{-2}
\end{equation}

\begin{equation}
\left(\frac{T_{0}}{\mathrm{K}}\right)  = 1.55 \times 10^{7} \left( \frac{v_{\mathrm{w}}}{10^{3} \mathrm{km\ s}^{-1}}\right)^{2}
\end{equation}

\begin{figure}[ht]
\begin{center}
\includegraphics[height=6.5cm, width=6.5cm]{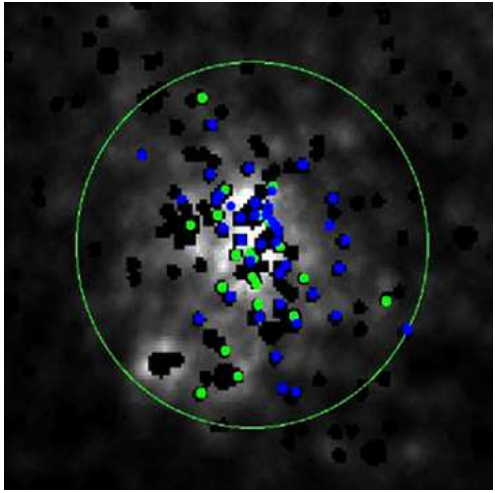}
\caption{Distribution of massive stars in the cluster core in relation to the diffuse emission. Wolf-Rayets are indicated by the filled green circles whereas main sequence and evolved OB stars are indicated by filled blue circles. The large circle indicates the 2'' core radius and the black areas indicate areas where point sources were masked. It is clear that the diffuse emission resembles the distribution of the massive stars in the cluster core.}
\label{dist}
\end{center}
\end{figure}

\par \noindent where $N$ is the number of stars contributing to the thermalized cluster wind, $\dot{M}$ is the average mass loss rate per star, $v_{\mathrm{w}}$ is a weighted average wind velocity of the stars and $R_{\mathrm{C}}$ is the radius of the region containing the stars. To enable direct comparison of results we apply a similar treatment as performed by MU06. In the analysis of MU06 only the WR stars are considered as the contributors to the cluster wind and from Equation 1 an emission measure ($K_{\mathrm{EM}} = \frac{4}{3}\pi R_{\mathrm{C}}^{3}n_{0}^{2}$) is calculated before using a standard thermal plasma model in XSPEC to extract fluxes. Our treatment differs slightly from that of MU06 in two ways. First, we set the value of $R_{\mathrm{C}}$ at 2 pc ($\approx$2' at 3.55 kpc) which is our core extraction region in the analysis above. This is an acceptable value as \citet{Canto2000} state that the value of $R_{\mathrm{C}}$ needs only be approximate to the distance from the cluster centre to the outermost star, i.e. - all the stars considered to contribute to the cluster wind are inside $R_{\mathrm{C}}$ (see also Figure \ref{dist}). Second, rather than assign typical values for $\dot{M}$ and $v_{\mathrm{w}}$ for WRs in general, we use the known spectral types of the WRs in the cluster core \citep{Crowther2006} along with general physical and wind properties of WRs \citep{Crowther2007} to estimate the more accurate mean values of $\dot{M} = 1.4 \times 10^{-5}$ M$_{\odot}$ yr$^{-1}$ and $v_{\mathrm{w}} = 1320 $\ km s$^{-1}$ for the 21 WRs within 2 pc of the cluster centre. Inputting these values into Equations 1 and 2 yields $n_{0} = 0.5$ cm$^{-3}$ and $kT_{0} = 3.7$ keV. We then calculate $K_{\mathrm{EM}}$ and feed the appropriate values into the APEC spectral model \citep{Smith2001} in XSPEC which gives an unabsorbed X-ray luminosity ($L_{\mathrm{X}}^{\mathrm{unabs}}$) in the 2-8 keV energy range of $\sim 2 \times 10^{33}$ erg\ s$^{-1}$. Extracting the observed high temperature value from our two temperature fit yields an $L_{\mathrm{X}}^{\mathrm{unabs}}$ of $1.7 \times 10^{33}$ erg\ s$^{-1}$. Hence, we find the predicted value is in excellent agreement with the observed value. Thus, a thermalized cluster wind can account for the hard thermal emission in the core of Wd1. Obviously, this remains true also if the estimated contribution by unresolved PMS objects (Section \ref{upms}) is subtracted from the observed hard emission.

This result differs from that of MU06, who found that the hard emission in the cluster core is approximately half of their predicted value. The discrepancy results mainly from the setting of $R_{\mathrm{C}} = 4\ \mathrm{pc}$ ($\approx$3' at their adopted distance of 5 kpc) in their calculations. Our smaller $R_{\mathrm{C}}$ value reduces the emission measure and hence the derived $L_{\mathrm{X}}^{\mathrm{unabs}}$. We note further that the morphology of the Wd1 diffuse X-ray emission (see Figure \ref{dist}) resembles the distribution of the massive stars in the cluster, in particular in showing a noticeable extension to the SE. When the less massive stars are considered, such as those down to $0.8 \mathrm{M}_{\odot}$ (Brandner et al. 2008; see also their Figure 1), the cluster shows a smoother, roughly N-S elongated distribution. Also these qualitative considerations link the diffuse emission with the massive stars (that exhibit winds) rather than the low-mass (PMS) objects.

\par \citet{Stevens2003} used the model of \citet{Canto2000} to predict the properties of the cluster wind for some Galactic and Magellanic Cloud clusters and compared them to observation. They found that, in all cases bar one, the predicted $kT_{0}$ values are much larger than observed and that the predicted X-ray luminosities are much less than observed. Our results are somewhat at odds with those of Stevens \& Hartwell in that our predicted values of $kT_{0}$ and $L_{\mathrm{X}}^{\mathrm{unabs}}$ are quite close to those observed. The difference in $kT_{0}$ can be explained by our adopted $v_{\mathrm{w}}$ value. For the determination of $kT_{0}$ we are only considering the WRs in the cluster (given that the early O stars have already evolved off the main sequence), which have lower wind velocities than the early O stars in the clusters studied by Stevens \& Hartwell, thus keeping $kT_{0}$ down. A possible explanation for their difference in $L_{\mathrm{X}}^{\mathrm{unabs}}$, and also $kT_{0}$ values, is that in all bar the same one cluster, Stevens \& Hartwell fit the diffuse spectra with a single temperature model. This may underestimate any cool thermal component that is present in the diffuse spectra of the clusters and the observed $kT$ value will be an amalgam of the hot and cool component $kT$ values (in fact, results for $kT$ and $L_{\mathrm{X}}^{\mathrm{unabs}}$ from single temperature fits will be coupled). If a cool component has been underestimated, this will dramatically affect the observationally derived $L_{\mathrm{X}}^{\mathrm{unabs}}$ given that the unabsorbed cool component will greatly add to the overall $L_{\mathrm{X}}^{\mathrm{unabs}}$. Since we allow for a cool thermal component in Wd1 and restrict our analysis to the harder thermal component only, our results are close to those predicted. One must be aware however that the model used in our analysis does not incorporate the thermalization efficiency and mass loading of the cluster wind described by Stevens \& Hartwell. In practice, our adopted model assumes no mass loading and a thermalization efficiency of 1 (i.e. -  no radiative losses in the conversion of the stellar wind energies to the cluster wind). This may be simplistic and we note here that a change in either parameter would serve to increase the predicted overall X-ray luminosity (i.e. - not restricted to the hard energy band) and reduce the $kT_{0}$ value.

\subsubsection{Supernova Remnants}
SNRs emit X-rays through thermal and/or non-thermal processes. Since we have identified the hard emission in the core of Wd1 as predominantly thermal, we address only the thermal emission mechanisms (i.e. -  the SNR interaction with its surroundings). The occurrence of SuperNovae (SNae) in Wd1 has been considered in previous analyses of Wd1 \citep[MU06; ][ etc.]{Clark2008,Brandner2008}, which use the stellar population to extrapolate the cluster IMF to higher masses and determine that the cluster initially contained $\sim10^{2}$ stars with $\mathrm{M} > 50 \mathrm{M_{\odot}}$. Given the age of Wd1 we can assume that all these stars have already been lost to SNae. However, apart from the magnetar, no evidence of post-SN objects (SNRs, compact objects or X-ray Binaries) was found in the analyses of either MU06 or \citet{Clark2008}. Possible reasons for the absence of the discrete objects are beyond the scope of this analysis, instead we only address the potential contribution of any SNR to the diffuse emission in the cluster core. SNRs are expected to emit X-rays when the shock front interacts with the surrounding ISM, but in the Wd1 region winds from the massive stars have cleared away the ISM so we cannot expect to observe emission from this process (MU06 and references therein).

However, SNRs can also interact with the stellar winds of nearby massive stars in the cluster. An illustrative example is the LBV/WR binary system HD 5980 in the Small Magellanic Cloud \citep{Naze2002}, the stellar wind of which is likely interacting with the ejecta from the nearby object SNR 0057-7226. \citet{Naze2002} performed an analysis of the diffuse emission around the HD 5980 system finding several X-ray bright filaments in the diffuse structure and obtaining a 0.3-10 keV $L_{\mathrm{X}}^{\mathrm{unabs}}$ of $\sim10^{35} \mathrm{erg\ s}^{-1}$, results that both were subsequently derived theoretically by \citet{Vel2003}. Additionally \citet{Naze2002} found that the diffuse emission associated with the object is best fit by an absorbed thermal plasma with $kT=0.66$ keV. This value is much lower than the hard component plasma temperature of the diffuse emission in Wd1, already suggesting that SNRs are not primarily responsible for the hard diffuse emission in the core since this plasma temperature cannot account for the observed 6.7 keV emission line. However, in view of the large luminosity found by \citet{Naze2002} for the diffuse emission around HD 5980, it is possible that the high energy tail of a SNR/stellar wind interaction spectrum may contribute to the observed hard diffuse emission in Wd1. To assess this we consider the cool thermal component of the Wd1 core diffuse spectrum which has a derived plasma temperature of 0.68 keV, similar to that of the HD 5980 diffuse emission. From the results of the spectral fits in Table \ref{diffuseparam} we estimate that the soft component contributes about one third of the 2-8 keV $L_{\mathrm{X}}^{\mathrm{unabs}}$ in the core. Or, if the estimated hard emission contribution by unresolved PMS objects of $0.46-1\times10^{33} \mathrm{erg\ s}^{-1}$ (Section \ref{upms}) is subtracted from the observed hard diffuse emission, about 40-60\% of the remaining hard emission could be contributed by the soft component. Thus even if SNRs were responsible for all of the observed soft emission, they would only account for between one third to one half of the observed hard emission in the core but crucially could not account for the 6.7 keV emission line.

In itself, any SNR contribution to diffuse emission is of course dependent on the recent occurrence of a SN event in or near the core. \citet{Muno2006} determine the SN rate in Wd1 to be once every 7,000-13,000 yr. If we assume that a SN event occurs at the very centre of the cluster core and that the SN ejecta travel with a velocities of a few $10^{3}$ km s$^{-1}$, then the ejecta escape the core region after several $10^{2}$ yr (and the cluster itself after a few $10^{3}$ yr). This leaves a significant amount of time (typically over 90\% of the time between successive SNae) in which the core is free from the effects of SNRs. Also this time scale argument suggests that it is unlikely that SNRs are contributing significantly to the Wd1 diffuse emission at the current epoch.

\section{Conclusions}

The above analysis and discussion of the until now unexploited XMM-Newton diffuse emission data for Wd1 have demonstrated a Fe 6.7 keV emission line, indicating that the hard X-ray component in the inner 2' radius region of the cluster is predominantly thermal in origin. The most likely explanation for this diffuse component is a thermalized cluster wind. An estimated value for the 2-8 keV X-ray luminosity produced by a cluster wind in Wd1 ($2 \times 10^{33} \mathrm{erg\ s}^{-1}$) is close to the observationally determined value for this luminosity ($1.7 \times 10^{33} \mathrm{erg\ s}^{-1}$). The conclusion that the cluster wind is the likely cause of the diffuse emission is also in line with the model predictions of \citet{Oskinova2005}.

While the unresolved PMS objects are less likely the main cause of the Wd1 diffuse emission, they could nevertheless still make a contribution to the diffuse component. We determined the unresolved PMS source contribution by comparing the observed PMS source flux distribution of Wd1 to that of the ONC. We found that the derived surface brightness value for the unresolved source population of $\sim4 \times 10^{31}$ erg s$^{-1}$ pc$^{-2}$ can only account for $\sim15$\% of the observed value of $\sim3 \times 10^{32}$ erg s$^{-1}$ pc$^{-2}$.

SNRs interacting with stellar winds are likely to be too soft X-ray emitters than is required to produce the 6.7 keV emission line. Individual SNRs have not been identified in Wd1 and, from time scale arguments, are a priori not likely to be present.

\section{Acknowledgements}
We wish to thank the anonymous referee for very constructive suggestions to improve the paper. This research was funded through an Irish Research Council for Science, Engineering and Technology Embark Initiative Scholarship awarded to Patrick Kavanagh.








\end{document}

%% file: aas_macros.tex
%
%
%


\def\jnl@style{\it}
\def\aaref@jnl#1{{\jnl@style#1}}

\def\aaref@jnl#1{{\jnl@style#1}}

\def\rspsa{\aaref@jnl{Royal~Soc.~of~Lon.~Proceed.~Ser.~A}}                   
\def\aj{\aaref@jnl{AJ}}                   
\def\araa{\aaref@jnl{ARA\&A}}             
\def\apj{\aaref@jnl{ApJ}}                 
\def\apjl{\aaref@jnl{ApJ}}                
\def\apjs{\aaref@jnl{ApJS}}               
\def\ao{\aaref@jnl{Appl.~Opt.}}           
\def\apss{\aaref@jnl{Ap\&SS}}             
\def\aap{\aaref@jnl{A\&A}}                
\def\aapr{\aaref@jnl{A\&A~Rev.}}          
\def\aaps{\aaref@jnl{A\&AS}}              
\def\azh{\aaref@jnl{AZh}}                 
\def\baas{\aaref@jnl{BAAS}}               
\def\jrasc{\aaref@jnl{JRASC}}             
\def\memras{\aaref@jnl{MmRAS}}            
\def\mnras{\aaref@jnl{MNRAS}}             
\def\pra{\aaref@jnl{Phys.~Rev.~A}}        
\def\prb{\aaref@jnl{Phys.~Rev.~B}}        
\def\prc{\aaref@jnl{Phys.~Rev.~C}}        
\def\prd{\aaref@jnl{Phys.~Rev.~D}}        
\def\pre{\aaref@jnl{Phys.~Rev.~E}}        
\def\prl{\aaref@jnl{Phys.~Rev.~Lett.}}    
\def\pasp{\aaref@jnl{PASP}}               
\def\pasj{\aaref@jnl{PASJ}}               
\def\qjras{\aaref@jnl{QJRAS}}             
\def\skytel{\aaref@jnl{S\&T}}             
\def\solphys{\aaref@jnl{Sol.~Phys.}}      
\def\sovast{\aaref@jnl{Soviet~Ast.}}      
\def\ssr{\aaref@jnl{Space~Sci.~Rev.}}     
\def\zap{\aaref@jnl{ZAp}}                 
\def\nat{\aaref@jnl{Nature}}              
\def\iaucirc{\aaref@jnl{IAU~Circ.}}       
\def\aplett{\aaref@jnl{Astrophys.~Lett.}} 
\def\apspr{\aaref@jnl{Astrophys.~Space~Phys.~Res.}}
\def\bain{\aaref@jnl{Bull.~Astron.~Inst.~Netherlands}} 
\def\fcp{\aaref@jnl{Fund.~Cosmic~Phys.}}  
\def\gca{\aaref@jnl{Geochim.~Cosmochim.~Acta}}   
\def\grl{\aaref@jnl{Geophys.~Res.~Lett.}} 
\def\jcp{\aaref@jnl{J.~Chem.~Phys.}}      
\def\jgr{\aaref@jnl{J.~Geophys.~Res.}}    
\def\jqsrt{\aaref@jnl{J.~Quant.~Spec.~Radiat.~Transf.}}
\def\memsai{\aaref@jnl{Mem.~Soc.~Astron.~Italiana}}
\def\nphysa{\aaref@jnl{Nucl.~Phys.~A}}   
\def\nphysb{\aaref@jnl{Nucl.~Phys.~B Proc. Supp.}}   
\def\physrep{\aaref@jnl{Phys.~Rep.}}   
\def\physscr{\aaref@jnl{Phys.~Scr}}   
\def\planss{\aaref@jnl{Planet.~Space~Sci.}}   
\def\procspie{\aaref@jnl{Proc.~SPIE}}   

\let\astap=\aap
\let\apjlett=\apjl
\let\apjsupp=\apjs
\let\applopt=\ao